\documentclass[10pt,letterpaper,twocolumn]{article} 

\usepackage{times,mathptmx}
\usepackage{ol2}
\usepackage{amsmath,amssymb}

\begin{document}

\twocolumn[ 

\title{Universal Scaling Laws of Kerr Frequency Combs}


\vskip-3mm

\author{St\'ephane Coen,$^{*}$ and Miro Erkintalo}

\address{
Physics Department, The University of Auckland, Private Bag 92019, Auckland 1142, New Zealand\\
$^*$Corresponding author: s.coen@auckland.ac.nz}

\begin{abstract}%
Using the known solutions of the Lugiato-Lefever equation, we derive universal trends of Kerr frequency combs. In
particular, normalized properties of temporal cavity soliton solutions lead us to a simple analytic estimate of
the maximum attainable bandwidth for given pump-resonator parameters. The result is validated via comparison with
past experiments encompassing a diverse range of resonator configurations and parameters.
\end{abstract}

\ocis{230.5750, 190.5530, 190.4380}

 ] 

\noindent Over the last few years, the generation of frequency combs in high-Q Kerr microresonators pumped by
continuous-wave (cw) laser light has attracted considerable interest \cite{kippenberg_microresonator-based_2011}.
Despite this interest, the dependence of comb characteristics on the pump-resonator parameter-space remains largely
unexplored, with little or no consensus existing on guidelines for Kerr comb optimization. This shortcoming
originates from the computational complexity of traditional models, such as the coupled-mode equation
model~\cite{chembo_modal_2010, matsko_hard_2012, matsko_chaotic_2013}. Approximate analytic solutions exist but the
comb characteristics still cannot be inferred in closed form~\cite{matsko_mode-locked_2011}.

We have recently shown that Kerr frequency combs can be efficiently modeled using a generalized mean-field
Lugiato-Lefever equation (LLE), and that they can be associated with the cavity soliton (CS) solutions of this
equation~\cite{coen_modeling_2013}. Also Herr et al. have presented convincing experimental evidence in strong
support of the CS hypothesis \cite{herr_soliton_2013}. Here, we use the mean-field framework to identify universal
trends in the dynamics and characteristics of Kerr frequency combs. Our objectives are two-fold: (i) we link the
known solutions of the LLE into the Kerr comb context and (ii) use the solutions to derive \textit{universal scaling
laws} that allow comb bandwidths to be analytically estimated. Comparison with past experiments across a wide
variety of resonator parameters and architectures shows good agreement with our results.

Our starting point is the normalized LLE~\cite{leo_temporal_2010, coen_modeling_2013},
\begin{equation}
  \label{LLN}
  \frac{\partial E(t,\tau)}{\partial t} = \left[ -1 +i(|E|^2- \Delta)
  -i\eta\frac{\partial^2}{\partial \tau^2}\right]E+S.
\end{equation}
Here $t$ is the slow time describing the evolution of the intracavity electric field envelope $E(t,\tau)$ at the
scale of the cavity photon lifetime while $\tau$ is a fast time defined in a reference frame traveling at the group
velocity of light in the resonator. The terms on the right-hand side of Eq.~\eqref{LLN} describe, respectively,
cavity losses, Kerr nonlinearity, pump-cavity detuning, 2nd-order group-velocity dispersion (GVD), and external
pumping. $\eta$ is the sign of the GVD coefficient $\beta_2$, and we assume here anomalous dispersion, $\eta=-1$.
The normalization follows Ref.~\cite{leo_temporal_2010}: ${t \rightarrow \alpha t/t_\mathrm{R}}$, ${\tau \rightarrow
\tau(2\alpha/|\beta_2|L)^{1/2}}$, and ${E \rightarrow E(\gamma L/\alpha)^{1/2}}$. Here $t_\mathrm{R} =
\mathrm{FSR}^{-1}$ is the cavity roundtrip time with FSR the free-spectral range, $\alpha =
(\alpha_\mathrm{i}L+\theta)/2$ is the total round-trip loss parameter with $\alpha_\mathrm{i}$ describing internal
linear absorption and $\theta$ the input coupler power transmission coefficient, and $L$ is the resonator length.
The nonlinearity coefficient $\gamma = 2\pi n_2/(\lambda_\mathrm{p} A_\mathrm{eff})$ with $n_2$ the nonlinear
refractive index, $\lambda_\mathrm{p}$ the pump wavelength, and $A_\mathrm{eff}$ the effective mode area. The
solutions of Eq.~\eqref{LLN} are governed by two normalized parameters: the pump strength $S = E_\mathrm{in} (\gamma
L \theta/\alpha^3)^{1/2}$, with $E_\mathrm{in}$ the cw pump amplitude in units of $W^{1/2}$, and the detuning
$\Delta = \delta_0/\alpha$, where $\delta_0$ is the phase detuning of the pump from the closest cavity resonance.
Note that $\Delta$ is formally identical to the $\zeta_0$ parameter introduced in Ref.~\cite{herr_soliton_2013}.

\begin{figure}[t]
  \includegraphics[clip=true]{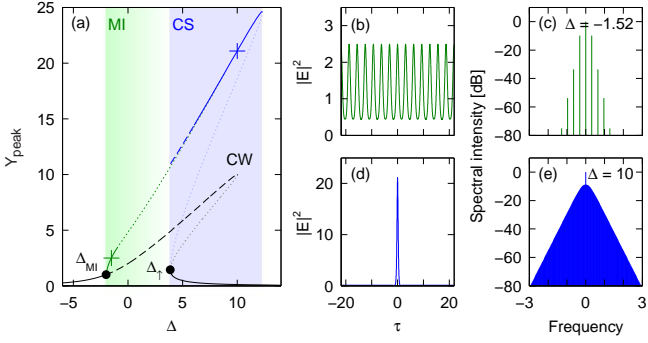}
  \vskip-2mm
  \caption{\small{(Color online) (a) Intracavity peak power versus detuning for the cw (black), MI (green), and CS (blue) branches. Dashed
    and dotted lines indicate unstable solutions as described in the text. The temporal and spectral intensity profiles of
    particular MI and CS solutions (crosses) are illustrated in (b,c) and (d,e), respectively.}}
  \label{fig1}
  \vskip-3mm
\end{figure}

Under cw pumping, the simplest steady-state solutions of Eq.~\eqref{LLN} are homogeneous ($\partial
E/\partial\tau=0$) and satisfy
\begin{equation}
  X = Y^3-2\Delta Y^2+(\Delta^2+1)Y,
  \label{bistability}
\end{equation}
where $X=|S|^2$ and $Y=|E|^2$ are the normalized pump and intracavity powers, respectively.
Equation~\eqref{bistability} is the well-known cubic equation of dispersive optical
bistability~\cite{haelterman_dissipative_1992}. At constant pump, it also describes the Kerr tilt of the cavity
resonances as shown in Fig.~\ref{fig1}(a) for $X = 10$ (black curve). Here $X$ is large enough for $Y(\Delta)$ to be
multivalued. The lower branch, existing for $\Delta>\Delta_\uparrow$, is always stable [$\Delta_\uparrow \simeq
3(X/4)^{1/3}$ for $X\gg1$] while the intermediate branch (dotted black) is homogeneously unstable. The left side of
the resonance, including the upper branch of the multivalued region, exhibits modulation instability (MI) for
intracavity powers above the threshold $Y>1$~\cite{haelterman_dissipative_1992}, or equivalently for detunings
$\Delta > \Delta_\mathrm{MI} = 1-\sqrt{X-1}$ (dashed black). Accordingly $X=1$ is the absolute MI threshold, also
referred to as the hyperparametric threshold in the comb literature~\cite{kippenberg_microresonator-based_2011,
chembo_modal_2010, matsko_hard_2012}.

MI leads to patterned solutions~\cite{haelterman_dissipative_1992} and the green curve in Fig.~\ref{fig1}(a)
represents the peak power of a range of such solutions ($X=10$ as above). They are steady-state periodic solutions
of Eq.~\eqref{LLN} and have been obtained with a Newton solver and a continuation method~\cite{coen_modeling_2013}.
Here the MI frequency is set to the most unstable one at threshold. For the point highlighted by the green cross, we
have also plotted in Figs.~\ref{fig1}(b)--(c) the corresponding temporal and spectral intensity profiles. As can be
seen, an MI solution corresponds in the spectral domain to a frequency comb, but resonators are typically long
enough to fit multiple periods of the pattern, so the comb modes would be separated by multiple FSRs. For larger
$\Delta$, the MI branch becomes unstable (dotted green), and split-step simulations of Eq.~\eqref{LLN} reveal a
transition to chaotic combs whose broad-linewidth components are separated by a single FSR, as also observed from
the coupled-mode equation model~\cite{matsko_chaotic_2013}.

CSs constitute the last class of steady-state solutions of Eq.~\eqref{LLN}~\cite{leo_temporal_2010,
coen_modeling_2013}. Their peak power, obtained as for the MI branch, completes Fig.~\ref{fig1}(a) [blue curve]. In
the time domain, CSs are localized pulses sitting atop a cw background and, due to resonator periodic boundary
conditions, correspond to a frequency comb in the spectral domain~\cite{coen_modeling_2013}. A typical solution
(blue cross) is illustrated in Figs.~\ref{fig1}(d)--(e). Here the comb separation is a single FSR as there is only
one CS in the cavity. The CS background matches with the lower state homogenous solution, as it is the only stable
cw solution, hence CSs only exist above the up-switching point, $\Delta>\Delta_\uparrow$. It is also found that CS
existence is limited to a maximum detuning of about $\pi^2 X/8$~\cite{barashenkov_existence_1996}. The lower part of
the CS branch is always unstable (dotted blue). On the upper branch, CSs go through a Hopf bifurcation for lower
detunings (dashed blue). Here CSs are oscillating (breathing) over multiple
round-trips~\cite{barashenkov_existence_1996, matsko_on-excitation_2012, leo_dynamics_2013}. More complex chaotic
regimes also exist in this region~\cite{leo_dynamics_2013}.

In the context of Kerr frequency combs, important conclusions emerge from the (co)existence of the cw, MI, and CS
solutions represented in Fig.~\ref{fig1}(a). CSs are the preferred solutions for combs because, in comparison to MI,
they are stable over a wider parameter range and their peak power is higher, leading to broader spectral bandwidths.
However, CSs cannot spontaneously emerge from an initial cw background as stable CS solutions are disconnected from
the cw branch~\cite{leo_temporal_2010}. Moreover, to exploit thermal self-locking, the resonance is approached from
the high frequency side [from the left in Fig.~\ref{fig1}(a)]. Combined with the fact that $\Delta_\mathrm{MI}$ is
\emph{always} smaller than $\Delta_\uparrow$, it implies that \textit{MI will always occur first}. The onset of MI
can therefore be associated with the ``primary combs'' described in other works~\cite{chembo_modal_2010}. The MI
branch extends into the CS branch, but unstable (chaotic or otherwise) MI and CS states lie in between the onset of
MI and the desirable stable CSs. Although beyond the scope of this Letter, preliminary simulations show that a
chaotic MI state can condensate into a set of CSs with appropriate ramping of~$\Delta$. Note that the chain stable
MI $\rightarrow$ chaotic MI $\rightarrow$ stable CSs as implied by Fig.~\ref{fig1}(a) is precisely that observed in
recent experiments \cite{herr_soliton_2013}.

\begin{figure}[t]
  \includegraphics[clip=true]{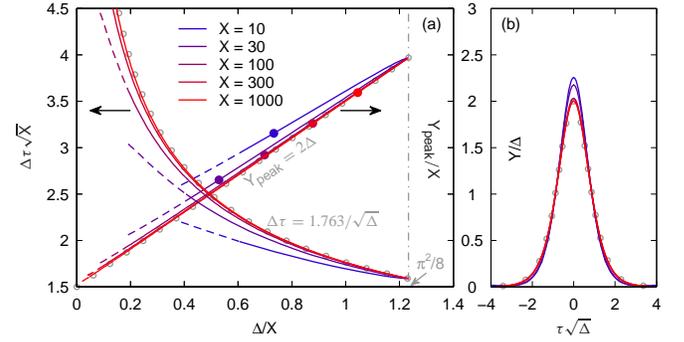}
  \vskip-2mm
  \caption{\small{(Color online) (a) Normalized CS duration (left axis) and peak power (right axis) versus $\Delta/X$ for various
    pump levels (dashed: oscillatory CSs; other unstable CSs are omitted for clarity). (b) Examples of numerical CS
    temporal intensity profiles for parameters corresponding to the dots
    in (a). In both panels, the numerical results are compared with analytic approximations (grey circles).}}
  \label{fig2}
    \vskip-2mm
\end{figure}

\begin{table*}[h]
\setlength{\tabcolsep}{5pt}
\caption{Comparison of experimental 3-dB comb bandwidths ($\Delta f_\mathrm{exp}$) with analytical predictions ($\Delta f_\mathrm{theo}$).}
\label{table}
\begin{center}
\begin{tabular}{cccccccccc}\hline
Ref. & $\gamma\,[\mathrm{W^{-1} km^{-1}}]$ & $P_\mathrm{in}$\,[mW] & $\lambda_\mathrm{p}$\,[nm] & Q & FSR\,[GHz] & $\mathcal{F}$ & $\beta_2\,[\mathrm{ps^2 km^{-1}}]$ & $\Delta f_\mathrm{exp}$\,[THz]  & $\Delta f_\mathrm{theo}$\,[THz]  \\ \hline
\cite{herr_soliton_2013} & 0.405 & 24   & 1553 & $400\cdot 10^6$  & 35.2 & $73 \cdot 10^3$ & $-5.9$ & 1.6 & 1.59 \\
\cite{dhaye_optical_2007} & 15 &  160 & 1550 & $100\cdot 10^6$  & 882 & $460 \cdot 10^3$ & $-6.3$ & 31 & 36.4 \\
\cite{okawachi_octave-spanning_2011}& 800 & 2000 & 1562 & $3\cdot 10^5$    & 226  & 350 & $-47$  & 10 & 15.6 \\
\cite{grudinin_frequency_2012} & 0.032 & 55.6 & 1560 & $1.90\cdot 10^9$ & 18.2 & $180 \cdot 10^3$ & $-13$  & 0.6 & 0.71 \\
\cite{dhaye_octave_2011} & 25 & 1000 & 1550 & $270 \cdot 10^6$ & 850 & $1.2 \cdot 10^6$ & $-4.0$ & 40 & 388 \\
\hline
\end{tabular}
\end{center}
  \vskip-5mm
\end{table*}

In the CS regime, the comb bandwidth is determined by the temporal duration of the CS. To gain some general
insights, we have calculated CS characteristics across a wide range of parameters. In Fig.~\ref{fig2}(a), we plot
the full-width at half-maximum (FWHM) $\Delta\tau$ of the CS solution (left axis) and its peak power
$Y_\mathrm{peak}$ (right axis) as a function of cavity detuning for several values of $X$. Note how we have
normalized all axes based on $X$ to reveal universal CS trends. We see that the CS peak power increases linearly
with detuning, with a slope approaching two for increasing values of $X$, i.e., $Y_\mathrm{peak} \simeq 2\Delta$. As
the peak power increases, the CS duration decreases and approaches $\Delta\tau \simeq 1.763/\sqrt{\Delta}$ for
$X\gg1$. In Fig.~\ref{fig2}(b) we also show normalized temporal intensity profiles of selected solutions [dots in
Fig.~\ref{fig2}(a)] superimposed with a $\mathrm{sech}^2$ pulse with peak power and duration as derived above.
Agreement is excellent for all parameters, implying that CSs can be approximated as $E_\mathrm{CS} \simeq
\sqrt{2\Delta}\,\mathrm{sech}(\sqrt{\Delta}\tau)$. It is worth noting that this expression is an exact analytical
solution of the normalized LLE \eqref{LLN} for a pulsed pump identical to the soliton, $S(\tau) = E_\mathrm{CS}$,
and is also found by perturbation theory as the fixed point of \eqref{LLN} \cite{wabnitz_suppression_1993,
herr_soliton_2013}, as well as in the solution derived by Barashenkov and Smirnov with $X = \Delta \gg 1$
\cite{barashenkov_existence_1996, matsko_mode-locked_2011}.

Considering $E_\mathrm{CS}$ to be a good approximation of the intracavity field allows us to derive a simple
estimate for the comb bandwidth obtainable for given pump-resonator parameters. Specifically, assuming on-resonance
pumping ($\Delta = X$) and critical coupling ($\alpha=\theta$), and transforming into physical units, we get the
following simple theoretical estimate for the 3-dB comb bandwidth,\\[-0.8ex]
\begin{equation}
  \label{est}
  \Delta f_\mathrm{theo}= \frac{0.315}{1.763}\sqrt{\frac{2\gamma P_\mathrm{in} Q \lambda_\mathrm{p} \mathrm{FSR}}{\pi c |\beta_2|}}
    = \frac{0.315}{1.763}\sqrt{\frac{2\gamma P_\mathrm{in} \mathcal{F}}{\pi |\beta_2|}}\,,
\end{equation}
with $P_\mathrm{in}=|E_\mathrm{in}|^2$ the pump power, $Q$ the quality factor of the cavity, and $c$ the speed of
light. Eq.~\eqref{est} also reveals the relevance of the various parameters. Recalling that the cavity finesse
$\mathcal{F} = \pi/\alpha = Q \lambda_\mathrm{p}\mathrm{FSR}/c$, it is clear that the suitability of a given
resonator for broadband comb generation is determined solely by its finesse (or losses), nonlinearity, and GVD
coefficient, and is independent of its FSR. Note that MI generally follows trends similar to CSs such that
Eq.~\eqref{est} is qualitatively correct irrespective of the operation regime.

\begin{figure}[b]
  \vspace*{-4mm}
  \centerline{\includegraphics[clip=true]{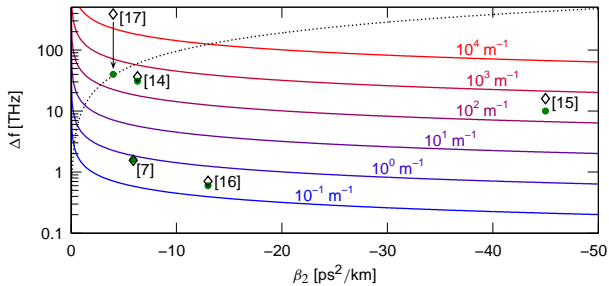}}
  \vskip-2.5mm
  \caption{\small{(Color online) Theoretical Kerr comb bandwidth calculated from Eq.~\eqref{est} versus 2nd-order GVD for varying
    values of $\gamma P_\mathrm{in} \mathcal{F}$ (solid lines) and upper limit (dotted line). The black diamonds
    are the theoretical estimates for the References listed in Table~\ref{table} with the corresponding experimental bandwidths shown as
    green dots.}}
  \label{fig3}
  \vskip -4mm
\end{figure}

The validity of Eq.~\eqref{est} has been checked by comparing its predictions against experimental comb bandwidths
for a number of references, as summarized in Table~\ref{table}, with all relevant experimental parameters also
listed. Aside from the comb reported in \cite{dhaye_octave_2011}, agreement is good throughout, with discrepancies
attributed to inaccuracies in experimental parameters, difficulties in estimating the bandwidths of
highly-structured experimental combs, as well as approximations implicit in Eq.~\eqref{est}. Note in particular how
the comparatively low finesse and high GVD of the resonator used in \cite{okawachi_octave-spanning_2011} is offset
by the large nonlinearity ($\gamma P_\mathrm{in}$ product). The large deviation between the estimated and
experimental bandwidths for \cite{dhaye_octave_2011} stems from the LLE~\eqref{LLN} and Eq.~\eqref{est} not taking
into account higher-order dispersion. This approximation becomes inaccurate for comb spectra extending into the
normal GVD regime, and results in significant over-estimation of the bandwidth. We can mitigate this issue by
recalling that CSs perturbed by higher-order dispersion emit dispersive waves (DW) into the normal GVD regime
\cite{erkintalo_cascaded_2012, coen_modeling_2013, leo_nonlinear_2013}. Assuming that the 3rd-order GVD coefficient
$\beta_3$ is the dominant contribution, the DW frequency shift can be written as $\Delta f_\mathrm{DW} =
3|\beta_2|/(2\pi \beta_3)$. Because the generated DWs are linear (they reside in a spectral region not supporting
solitons), no significant spectral broadening is expected beyond their spectral shift. We can thus interpret $\Delta
f_\mathrm{lim} = 2 \Delta f_\mathrm{DW}$ as being an \emph{upper limit} of the full-width bandwidth of a Kerr comb.
Interestingly, while Eq.~\eqref{est} suggests that low GVD is beneficial for broadband combs, the full-width limit
bandwidth $\Delta f_\mathrm{lim}$ is in this case constrained to a lower value. It is only a low dispersion
slope~$\beta_3$ (i.e., flat dispersion) that can lift this limit. These trends are illustrated in Fig.~\ref{fig3}
where we plot the bandwidths estimated from Eq.~\eqref{est} as a function of $\beta_2$ for various values of the
product $\gamma P_\mathrm{in} \mathcal{F}$ (solid lines) as well as the full-width upper limit $\Delta
f_\mathrm{lim}$ (dotted line), assuming here $\beta_3 = 0.1\ \mathrm{ps^3/km}$. In Fig.~\ref{fig3} we also show the
experimental (green dots) and theoretical [Eq.~\eqref{est}; black diamonds] bandwidths extracted from
Table~\ref{table}. Note in particular how the comb bandwidth observed in \cite{dhaye_octave_2011}, overestimated by
Eq.~\eqref{est}, falls on the theoretical upper limit.

To conclude, we have analyzed the solutions of a dimensionless mean-field equation from the Kerr comb perspective.
By identifying universal trends in the solutions we have derived universal scaling laws linking experimental
pump-resonator parameters to the obtainable comb bandwidth.

We acknowledge support from the Marsden Fund (Royal Society of New Zealand) and useful discussions with T. Herr.

\section*{References with titles}


\small

\begin{thebibliography}{10}

\bibitem{kippenberg_microresonator-based_2011}
T.~J. Kippenberg, R. Holzwarth, and S.~A. Diddams, ``Microresonator-based optical frequency combs,''
  Science {\bf 332}, 555--559 (2011).

\bibitem{chembo_modal_2010}
Y.~K. Chembo and N. Yu, ``Modal expansion approach to optical-frequency-comb generation with monolithic
  whispering-gallery-mode resonators,'' Phys. Rev. A {\bf 82},  033801/1--18  (2010).

\bibitem{matsko_hard_2012}
A.~B. Matsko, A.~A. Savchenkov, V.~S. Ilchenko, D. Seidel, and L. Maleki,
  ``Hard and soft excitation regimes of Kerr frequency combs,'' Phys.
  Rev. A {\bf 85},  023830/1--5  (2012).

\bibitem{matsko_chaotic_2013}
A. B. Matsko, W. Liang, A. A. Savchenkov, and L. Maleki, ``Chaotic dynamics of frequency combs generated
  with continuously pumped nonlinear microresonators,'' Opt. Lett. {\bf 38}, 525--527
  (2013).

\bibitem{matsko_mode-locked_2011}
A.~B. Matsko, A.~A. Savchenkov, W. Liang, V.~S. Ilchenko, D. Seidel, and L.
  Maleki, ``Mode-locked Kerr frequency combs,'' Opt. Lett. {\bf 36},  2845--2847  (2011).

\bibitem{coen_modeling_2013}
S. Coen, H. G. Randle, T. Sylvestre, and M. Erkintalo, ``Modeling of octave-spanning Kerr
  frequency combs using a generalized mean-field Lugiato–Lefever model,'' Opt. Lett. {\bf 38},  37--39
  (2013).

\bibitem{herr_soliton_2013}
T. Herr, V. Brasch, M. L. Gorodetsky, T. J. Kippenberg, ``Soliton mode-locking in optical
  microresonators,'' arXiv:1211.0733.

\bibitem{leo_temporal_2010}
F. Leo, S. Coen, P. Kockaert, S.-P. Gorza, Ph. Emplit, and M. Haelterman, ``Temporal cavity solitons
  in one-dimensional Kerr media as bits in an all-optical buffer,'' Nat.
  Photon. {\bf 4},  471--476  (2010).

\bibitem{haelterman_dissipative_1992}
M. Haelterman, S. Trillo, and S. Wabnitz, ``Dissipative modulation instability
  in a nonlinear dispersive ring cavity,'' Opt. Comm. {\bf 91,}
  401--407 (1992).

\bibitem{barashenkov_existence_1996}
I. V. Barashenkov and Yu. S. Smirnov, ``Existence and stability chart for the ac-driven, damped
  nonlinear Schr\"odinger solitons,'' Phys. Rev. A {\bf 54},  5707--5725  (1996).

\bibitem{matsko_on-excitation_2012}
A.~B. Matsko, A.~A. Savchenkov, and L. Maleki, ``On excitation of breather
  solitons in an optical microresonator,'' Opt. Lett. {\bf 37,} 4856--4858 (2012).

\bibitem{leo_dynamics_2013}
F. Leo, L. Gelens, Ph. Emplit, M. Haelterman, and S. Coen, ``Dynamics of one-dimensional Kerr
  cavity solitons,'' accepted in Opt. Express (2013).

\bibitem{wabnitz_suppression_1993}
S. Wabnitz, ``Suppression of interactions in a phase-locked soliton optical memory,''
  Opt. Lett. {\bf 18},  601--603  (1993).

\bibitem{dhaye_optical_2007}
P. {Del'Haye}, A. Schliesser, O. Arcizet, T. Wilken, R. Holzwarth, and T.~J.
  Kippenberg, ``Optical frequency comb generation from a monolithic microresonator,''
  Nature {\bf 450},  1214--1217 (2007).

\bibitem{okawachi_octave-spanning_2011}
Y. Okawachi, K. Saha, J.~S. Levy, Y.~H. Wen, M. Lipson, and A.~L. Gaeta,
  ``Octave-spanning frequency comb generation in a silicon nitride chip,'' Opt.
  Lett. {\bf 36},  3398--3400  (2011).

\bibitem{grudinin_frequency_2012}
I.~S. Grudinin, L. Baumgartel, and N. Yu, ``Frequency comb from a microresonator
  with engineered spectrum,'' Opt. Express {\bf 20},  6604--6609  (2012).

\bibitem{dhaye_octave_2011}
P. {Del'Haye}, T. Herr, E. Gavartin, M. L. Gorodetsky, R. Holzwarth, and T.~J.
  Kippenberg, ``Octave spanning tunable frequency comb from a microresonator,''
  Phys. Rev. Lett. {\bf 107}, 063901/1--4 (2011).

\bibitem{erkintalo_cascaded_2012}
M. Erkintalo, Y.~Q. Xu, S.~G. Murdoch, J.~M. Dudley, and G. Genty, ``Cascaded phase
  matching and nonlinear symmetry breaking in fiber frequency combs,'' Phys. Rev.
  Lett. {\bf 109},  223904/1--5  (2012).

\bibitem{leo_nonlinear_2013}
F. Leo, A. Mussot, P. Kockaert, Ph. Emplit, M. Haelterman, and M. Taki,
  ``Nonlinear symmetry breaking induced by third-order dispersion in optical fiber
  cavities,'' Phys. Rev. Lett. {\bf 110,} 104103/1--5 (2013).

\end{thebibliography}
\end{document}